\documentclass[aps,nofootinbib,prd,preprint,longbibliography,preprintnumbers,showpacs]{revtex4-1}
 \usepackage[utf8]{inputenc}
 \usepackage[T1]{fontenc} 
 \usepackage{amsmath}
 \usepackage{amsfonts}
 \usepackage{amssymb}
 \usepackage{amsthm}
 \usepackage{array} 
 \usepackage{booktabs} 
 \usepackage[margin=3cm]{geometry} 
 \usepackage{footnote}
 \usepackage{mathtools}
 \usepackage[pdftex]{hyperref} 
 \usepackage{cancel} 
 \usepackage{graphicx}
 \usepackage{float} 
 \usepackage{subcaption} 
 \usepackage{color}
\newcommand{\bra}[1]{\langle #1|} 
\newcommand{\ket}[1]{|#1\rangle}

 \date{\today}

\begin{document}
\raggedbottom
\preprint{HIP-2017-29/TH} 
 \title{The effects of triplet Higgs bosons in long baseline neutrino experiments}

\author{K. Huitu\footnote{e-mail: katri.huitu@helsinki.fi}}
\affiliation{Department of Physics, and Helsinki Institute of Physics, P. O. Box 64, FI-00014 University of Helsinki, Finland}
\author{T. J. Kärkkäinen\footnote{e-mail: timo.j.karkkainen@helsinki.fi}}
\affiliation{Department of Physics, and Helsinki Institute of Physics, P. O. Box 64, FI-00014 University of Helsinki, Finland}
\author{J. Maalampi\footnote{e-mail: jukka.maalampi@jyu.fi} and S. Vihonen\footnote{e-mail: sampsa.p.vihonen@student.jyu.fi}}
\affiliation{University of Jyvaskyla, Department of Physics, P.O. Box 35, FI-40014 University of Jyvaskyla, Finland}

\begin{abstract}
The triplet scalars $(\Delta=\Delta^{++},\Delta^{+},\Delta^{0})$, utilized in the so-called Type-II seesaw model to explain the lightness of neutrinos, would generate nonstandard interactions (NSI) for neutrino propagating in matter. We investigate the prospects to probe these interactions in long baseline neutrino oscillation experiments. We analyze the upper bounds that the proposed DUNE experiment might set on the nonstandard parameters and numerically derive upper bounds, as function of the lightest neutrino mass, on the ratio the mass $M_\Delta$ of the triplet scalars and strength $|\lambda_\phi|$ of  the coupling  $\phi\phi\Delta$ of the triplet $\Delta$ and conventional Higgs doublet $\phi$ . We also discuss the possible misinterpretation of these effects as effects arising from a nonunitarity of the neutrino mixing matrix and compare the results with the bounds that arise from the charged lepton flavor violating processes.

\end{abstract}

\maketitle

\section{Introduction}
The discovery of neutrino oscillations in atmospheric and solar neutrino measurements \cite{Kajita:2000mr,Ahmad:2001an} proved that the SU(2)$_{\rm L}\times$U(1)$_{\rm Y}$ Standard Model (SM) is not capable to fully explain particle physics world. The existence of oscillations indicates that neutrinos are massive particles, in contrast with the prediction of the SM. One has to go beyond the SM in order to discover the origin of neutrino masses. At the same time, one has to find a convincing explanation for the lightness of neutrinos as compared with the other basic particles, i.e. quarks and charged leptons. The most popular answer to the latter question is the so-called seesaw mechanism \cite{Minkowski:1977sc,GellMann:1980vs,Yanagida:1979as,Mohapatra:1979ia,Schechter:1980gr}, where the suppression of neutrino masses follows from the existence of a new mass scale much higher than the electroweak scale ${\cal O}(10^2)$ GeV. In the Type-I seesaw mechanism the new mass scale is set by sterile right-handed neutrinos by which the particle content of the SM is extended. The Type-II seesaw mechanism is based on the existence of a set of new scalars $\Delta =(\Delta^{++},\Delta^{+},\Delta^{0}$) transforming as a triplet under the SU(2)$_{\rm L}$ gauge symmetry. The masses of neutrinos are proportional to the ratio $\lambda_{\phi}v^2/M_{\Delta}^2$, where $v\simeq 174$ GeV is the vacuum expectation value of the SM Higgs field $\phi$, $M_{\Delta}$ is the mass of triplet scalar, and $\lambda_{\phi}$ is the dimensionful strength of the $\phi\phi\Delta$ coupling. 
 
 While the seesaw mechanism itself cannot be experimentally verified, the extension of the SM it is based on generally leads to experimentally testable phenomena. For example, the doubly charged scalar $\Delta^{++}$  would have clear experimental signatures e.g. in high-energy proton-proton collision experiments \cite{Huitu:1996su,Maalampi:2002vx,Muhlleitner:2003me,Akeroyd:2005gt,Bambhaniya:2015wna,ATLAS:2014kca}. One theoretical framework where a scalar triplet, as well as right-handed neutrinos, naturally appear is the left-right symmetric electroweak model based on the gauge symmetry SU(3)$_\text{C}\times$SU(2)$_\text{L}\times$SU(2)$_\text{R}\times$U(1)$_{B-L}$ \cite{Pati:1974yy,Mohapatra:1974gc,Mohapatra:1974hk,Senjanovic:1975rk,Gunion:1989in}. 
 
 In this paper we will concentrate on the Type-II seesaw mechanism and investigate how the triplet scalar bosons $\Delta$ would affect neutrino propagation in matter (for earlier studies, see e.g. \cite{Malinsky:2008qn}) and how these effects could be probed in long baseline neutrino experiments, particularly in the planned DUNE. Applying the bounds derived for DUNE in Ref. \cite{Blennow:2016etl}, together with the constraints one has for the elements of the neutrino mass matrix from earlier oscillation experiments, we compute an upper limit of the ratio $M_\Delta/|\lambda_\phi|$ as a function of the absolute neutrino mass scale (the mass $m_1$ of the lightest neutrino). For comparison, we also compute the upper bound for this ratio using the existing constraints on the charged lepton flavor violation (CLFV) processes.

\section{Nonstandard interactions and neutrino masses in a triplet model}
Our theoretical framework is the SU(2)$_{\rm L}\times$U(1)$_{\rm Y}$ electroweak model added with scalar triplet field $\Delta =(\Delta_1,\Delta_2,\Delta_3)\sim ({\bf 3},2)$, which can also be understood as a low-energy effective theory of the left-right symmetric SU(3)$_C\times$SU(2)$_L\times$SU(2)$_R\times$U(1)$_{B-L}$ theory where all the other nonstandard degrees of freedom except the triplet scalar are so heavy that they do not have observable effects in the oscillation experiments. The interactions of the triplet $\Delta$ relevant for the neutrino oscillation are described with the following Lagrangian:
\begin{equation}\label{eq:x:01}
\mathcal{L}_\Delta = Y_{\alpha\beta} \, L_{\alpha L}^{T} \, C \, i\sigma_2 \, \Delta \, L_{\beta L}+ \lambda_\phi \, \phi^T \, i\sigma_2 \, \Delta^\dagger \phi + \text{h.c.},
\end{equation}
 where $Y_{\alpha\beta}$ ($\alpha,\beta=e,\mu,\tau$) are Yukawa coupling constants, $C$ is the charge conjugation operator, $\phi$ is the SM Higgs doublet and the triplet $\Delta$ is presented in the  $2\times 2$  matrix form
 \begin{equation}\label{eq:x:02}
 \Delta =\frac{1}{\sqrt{2}}\sigma_i\Delta_i=
 \left( \begin{array}{cc}
\frac{\displaystyle{\Delta^{+}}}{\displaystyle\sqrt{2}}& \Delta^{++}\\
\Delta^0& - \frac{\displaystyle{\Delta^{+}}}{\displaystyle\sqrt{2}}
\end{array}\right),
\end{equation}
 where $\sigma_i$ are the Pauli matrices. When written in terms of component fields, Eq. (\ref{eq:x:01}) takes the form
\begin{equation}\label{eq:x:03}
\mathcal{L}_Y = Y_{\alpha\beta} \, \left[ \Delta^0 \, \overline{\nu_{\alpha R}^C}\, \nu_{\beta L} - \frac{1}{\sqrt{2}} \Delta^+ \, \left(\overline{\ell_{\alpha R}^C} \, \nu_{\beta L} + \overline{\nu_{\alpha R}^C} \, \ell_{\beta L} \right) - \Delta^{++} \, \overline{\ell_{\alpha R}^C} \, \ell_{\beta L}\right] + \text{h.c.}
\end{equation}
 These interactions lead in the second order of perturbation theory to the four-fermion interactions presented in the Fig. 1. The amplitude presented in Fig. 1(a) gives rise to Majorana mass terms for the neutrinos when the SU(2)$_{\rm L}\times$U(1)$_{\rm Y}$ symmetry is spontaneously broken, while the amplitudes in Fig. 1(b) and Fig. 1(c) correspond to new, nonstandard interactions among leptons. In the limit, where the mass of the triplet scalars $M_{\Delta}$, assumed to be the same for all members of the triplet, is large compared with the momenta of the processes, the amplitudes are described by the following effective Lagrangians \cite{Malinsky:2008qn}:
 \begin{equation}\label{eq:2}
\mathcal{L}_\nu^m = \frac{Y_{\alpha\beta} \, \lambda_\phi \, v^2}{M_\Delta^2} \,\, \left( \overline{\nu_{\alpha R}^C}  \, \nu_{\beta L } \right) = -\frac{1}{2} \, (m_\nu)_{\alpha\beta} \,  \overline{\nu_{\alpha R}^C}\, \nu_{\beta L },
\end{equation}

\begin{equation}\label{eq:3}
\mathcal{L}_\text{NSI} = \frac{Y_{\sigma\beta} \, Y_{\alpha\rho}^\dagger}{M_\Delta^2} \,\, \left( \overline{{\nu}_{\alpha L}} \, \gamma_\mu  \, \nu_{\beta L} \right) \,\, \left( \overline{{\ell}_{\rho L}}\, \gamma^\mu \, \ell_{\sigma  L} \right),
\end{equation}
where $m_\nu$ is the neutrino mass matrix, $M_\Delta$ is the degenerate mass of the $\Delta$ particles, and \textit{v} is the vacuum expectation value of the SM scalar Higgs field. The connection to the effective field theory can be derived by solving the Yukawa coupling $Y_{\alpha\beta}$ from the Majorana mass term in Eq. (\ref{eq:2}) and inserting it to the neutrino matter NSI term in Eq. (\ref{eq:3}). Comparing the result with the effective NSI Lagrangian 
\begin{equation}\label{eq:4}
\mathcal{L}_\text{NSI} = -2\sqrt{2}G_F\varepsilon^{ff' C}_{\alpha\beta}(\overline{\nu}_{\alpha L}\gamma^{\mu}\nu_{\beta L})(\overline{f}\gamma_{\mu}P_Cf'),
\end{equation}
where $P_C$ is chiral projection operator ($C=L,R$), $G_F$ is Fermi coupling constant, $f,f'$ are any fermions, $\alpha, \beta = e,\mu ,\tau$, and allowing only left-handed lepton terms (since $\Delta$ is leptophilic) one obtains the following expression for the nonstandard interaction parameters:
\begin{equation}\label{eq:5}
\varepsilon_{\alpha\beta}^{\rho\sigma} = -\frac{M_\Delta^2}{8\sqrt{2}\,G_F\,v^4\,\lambda_\phi^2} \,\, (m_\nu)_{\sigma\beta} \,\, (m_\nu^\dagger)_{\alpha\rho},
\end{equation}
where $\alpha,\beta,\rho$ and $\sigma$ are flavor indices. The expression (\ref{eq:5}) indicates the larger the ratio $M_\Delta^2/\lambda_\phi^2$ the stronger are nonstandard interactions of light neutrinos. Conversely, stricter bounds on $\varepsilon_{\alpha\beta}^{\rho\sigma}$ also mean better constraints on $M_\Delta^2/\lambda_\phi^2$.

\begin{figure}[H]
\begin{subfigure}{0.41\textwidth}
\includegraphics[width=\linewidth]{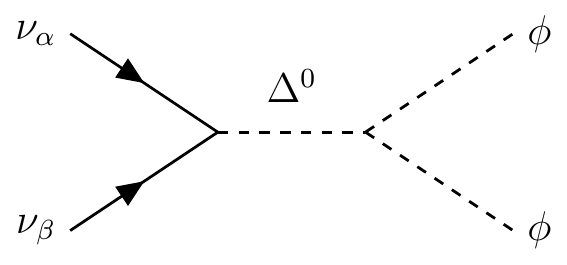}
\caption{Light neutrino Majorana mass term} \label{fig:a}
\end{subfigure}\hspace*{\fill}
\begin{subfigure}{0.41\textwidth}
\includegraphics[width=\linewidth]{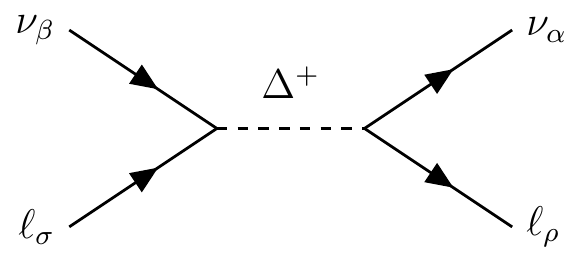}
\caption{Light neutrino matter NSI} \label{fig:b}
\end{subfigure}
\medskip
\begin{subfigure}{0.41\textwidth}
\includegraphics[width=\linewidth]{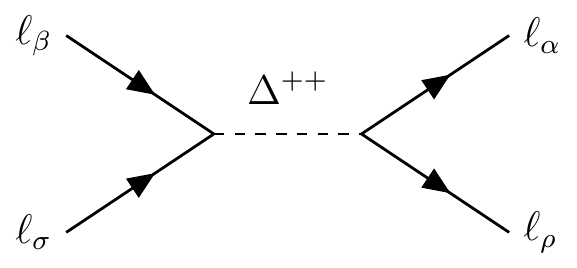}
\caption{Four-lepton NSI} \label{fig:c}
\end{subfigure}\hspace*{\fill}
\begin{subfigure}{0.41\textwidth}
\includegraphics[width=\linewidth]{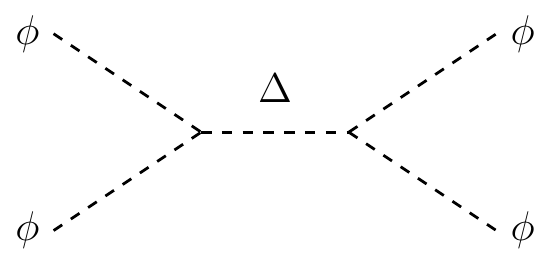}
\caption{SM Higgs self-coupling} \label{fig:d}
\end{subfigure}
\caption{\label{fig:1}Tree level Feynman diagrams for interactions between neutrinos $\nu$, leptons $\ell$ and the Standard Model Higgs scalar $\phi$, and are mediated by the triplet Higgs fields $\Delta$.}
\end{figure}

\section{\label{sec:3}Nonstandard interactions and nonunitary mixing}

In the low energy limit, nonstandard interactions stem from the effective charged current (CC) and neutral current (NC) like Lagrangians, which are given by \cite{Grossman:1995wx}:
\begin{equation}\label{lnsi}
\begin{split}
&\mathcal{L}^\text{CC}_\text{NSI} = -2\sqrt{2}G_F\varepsilon_{\alpha\beta}^{ff',C}(\overline{\nu}_{L\alpha}\gamma^{\mu}\ell_{L\beta})(\overline{f}\gamma^{\mu}P_Cf'),\\
&\mathcal{L}^\text{NC}_\text{NSI} = -2\sqrt{2}G_F\varepsilon_{\alpha\beta}^{f,C}(\overline{\nu}_{L\alpha}\gamma^{\mu}\nu_{L\beta})(\overline{f}\gamma^{\mu}P_Cf).
\end{split}
\end{equation}

 The CC Lagrangian $\mathcal{L}^\text{CC}_\text{NSI}$ in Eq. (\ref{lnsi}) is responsible for the NSI effects involved in neutrino source and detector, where nonstandard effects involve charged fermions. The NC Lagrangian $\mathcal{L}^\text{NC}_\text{NSI}$, on the other hand, is relevant for neutrino propagation in matter.

 Most of the charged current NSI parameters have been studied comprehensively in lepton decay experiments resulting in strict constraints. The current bounds on the CC NSI parameters $\varepsilon_{\alpha\beta}^{ff',C}$, as quoted in Ref. \cite{Malinsky:2008qn}, are presented in Table \ref{bounds:1}.

 In the future long baseline experiments, the NSI effects become particularly relevant in the neutrino propagation in matter, where they are covered by the effective NC Lagrangian $\mathcal{L}^\text{NC}_\text{NSI}$. For previous studies, see \cite{Huitu:2016bmb, Blennow:2016etl, Blennow:2016jkn, Ohlsson:2012kf, Girardi:2014kca, Meloni:2009ia, Biggio:2009nt, Kopp:2007ne, Adhikari:2012wn, Coloma:2015kiu, deGouvea:2015ndi, Mehta:2015, Fukasawa:2016lew, Davidson:2003ha, Ge2016}. The effective Hamiltonian takes the form
\begin{equation}
H=\frac{1}{2E_{\nu}}\left [U{\rm diag}(m_1^2,m_2^2,m_3^2)U^{\dagger} +{\rm diag}(A,0,0) + A\varepsilon^m \right],
\end{equation}
where $E_\nu$ is the energy of the propagating neutrino, \textit{U} is the light neutrino mixing matrix and $m_1$, $m_2$ and $m_3$ are the three masses of the active neutrinos. 

 In this formalism, the so-called matter NSI effects are parametrized as
\begin{equation}
\varepsilon^m_{\alpha\beta}=\sum_{f,C} \varepsilon_{\alpha\beta}^{f,C}\frac{N_f}{N_e},
\end{equation}
where $\varepsilon_{\alpha\beta}^{f,C}$ are the NSI parameters from the low energy NC Lagrangian of Eq. (\ref{lnsi}) and $N_f/N_e$ is the fraction of fermions of flavor \textit{f} over electrons in the medium the neutrino traverses.

 The effective matter potential is given by the matrix
\begin{equation}
V = A\left(
\begin{array}{ccc}
1+\varepsilon_{ee}^m & \varepsilon_{e\mu}^m & \,\,\, \varepsilon_{e\tau}^m \\
\varepsilon_{e\mu}^{m*} & \varepsilon_{\mu\mu}^m & \,\,\, \varepsilon_{\mu\tau}^m \\
\varepsilon_{e\tau}^{m*} & \varepsilon_{\mu\tau}^{m*} & \,\,\, \varepsilon_{\tau\tau}^m
\end{array}\right),
\label{Vnsi}
\end{equation}
where $A = \sqrt{2}G_FN_e$, $G_F$ is Fermi coupling constant and $N_e$ is the electron number density of the medium. The matter NSI effects are incorporated in the SM matter effects via parameters $\varepsilon_{\alpha\beta}^m$, $\alpha,\beta=e,\mu,\tau$.

 The effective Hamiltonian can therefore be written as
\begin{equation}
H = \frac{1}{2E_{\nu}}\left[U
\left(
\begin{array}{ccc}
0 \,\,\, & 0 & 0 \\
0 \,\,\, & \Delta m_{21}^2 & 0\\
0 \,\,\, & 0 & \Delta m_{31}^2 
\end{array}
\right) U^{\dagger}
+V\right],
\label{Hnsi}
\end{equation}
and the probability for the transition $\nu_\alpha \rightarrow \nu_\beta$ is given by
\begin{equation}
P_{\nu_{\alpha} \rightarrow \nu_{\beta}} = \left| \bra{\nu_{\beta}^{d}} e^{-iHL} \ket{\nu_{\alpha}^{s}}\right|^2,
\label{transprob}
\end{equation}
where $\ket{\nu_\alpha^s} \equiv \ket{\nu_\alpha} + \varepsilon_{\alpha \beta}^s \ket{\nu_\beta}$ and $\bra{\nu_\beta^d} \equiv \bra{\nu_\beta} + \varepsilon_{\alpha \beta}^d \bra{\nu_\alpha}$ refer to a neutrino that starts in flavor $\alpha$ in the source and is detected in flavor $\beta$ in the detector after the traversing distance \textit{L}. In long baseline experiments \textit{L} is set by the baseline of the experiment and $E_\nu$ by the energy of the neutrino beam. The matrices $\varepsilon^s$ and $\varepsilon^d$ contain dimensionless elements, which slightly alter the neutrino production and detection interactions in the source and detector, respectively. They originate from NSI-CC parameters $\varepsilon^{ff'C}_{\alpha\beta}$ in Eq. (\ref{lnsi}) and their current experimental bounds are reported in Ref.~\cite{Blennow:2016etl}.

\begin{table}[t]
\caption{\label{bounds:1}Constraints on the parameters $\varepsilon^{\rho\sigma}_{\alpha\beta}$ (updated from \cite{Malinsky:2008qn} using \cite{Patrignani:2016xqp}) from the $\ell \rightarrow \ell\,\ell\,\ell$, one-loop $\ell \rightarrow \ell\,\gamma$ and $\mu^+ e^- \rightarrow \mu^- e^+$ processes. All bounds are given at 90\% confidence level.}
\begin{center}
\begin{tabular}{|c|c|c|}\hline
\rule{0pt}{3ex}\textbf{Decay} & \textbf{Constraint on} & \textbf{Bound}\\ \hline
\rule{0pt}{3ex}$\mu^- \rightarrow e^- e^+ e^-$ & $|\varepsilon^{e\mu}_{ee}|$ & $3.5 \times 10^{-7}$ \\ \hline
\rule{0pt}{3ex}$\tau^- \rightarrow e^- e^+ e^-$ & $|\varepsilon^{e\tau}_{ee}|$ & $1.4 \times 10^{-4}$ \\ \hline
\rule{0pt}{3ex}$\tau^- \rightarrow \mu^- \mu^+ \mu^-$ & $|\varepsilon^{\mu\tau}_{\mu\mu}|$ & $1.2 \times 10^{-4}$ \\ \hline
\rule{0pt}{3ex}$\tau^- \rightarrow e^- \mu^+ e^-$ & $|\varepsilon^{e\tau}_{e\mu}|$ & $1.0 \times 10^{-4}$ \\ \hline
\rule{0pt}{3ex}$\tau^- \rightarrow \mu^- e^+ \mu^-$ & $|\varepsilon^{\mu\tau}_{\mu e}|$ & $1.0 \times 10^{-4}$ \\ \hline
\rule{0pt}{3ex}$\tau^- \rightarrow e^- \mu^+ \mu^-$ & $|\varepsilon^{e\tau}_{\mu\mu}|$ & $1.0 \times 10^{-4}$ \\ \hline
\rule{0pt}{3ex}$\tau^- \rightarrow e^- e^+ \mu^-$ & $|\varepsilon^{e\tau}_{\mu e}|$ & $9.9 \times 10^{-5}$ \\ \hline
\rule{0pt}{3ex}$\mu^- \rightarrow e^- \gamma$ & $|\sum_\alpha{\varepsilon^{e\mu}_{\alpha\alpha}}|$ & $2.6 \times 10^{-5}$ \\ \hline
\rule{0pt}{3ex}$\tau^- \rightarrow e^- \gamma$ & $|\sum_\alpha{\varepsilon^{e\tau}_{\alpha\alpha}}|$ & $1.8 \times 10^{-2}$ \\ \hline
\rule{0pt}{3ex}$\tau^- \rightarrow \mu^- \gamma$ & $|\sum_\alpha{\varepsilon^{\mu\tau}_{\alpha\alpha}}|$ & $2.0 \times 10^{-4}$ \\ \hline
\rule{0pt}{3ex}$\mu^+ e^- \rightarrow \mu^- e^+$ & $|\varepsilon^{\mu e}_{\mu e}|$ & $3.0 \times 10^{-3}$ \\ \hline
\end{tabular}
\end{center}
\end{table}

 It is possible that a part of the experimental signal presumed to originate from the matter NSI effects is actually caused by the nonunitarity of the light neutrino mixing matrix (see e.g. \cite{Escrihuela:2015wra,Escrihuela:2016ube}). We shall evaluate in our numerical analysis how large this contribution could possibly be in the triplet model, Eq. \eqref{eq:x:01}. In the case of nonunitarity, the light neutrino mixing matrix $U$ in Eq. (\ref{Hnsi}) must be replaced with a nonunitary matrix, which we denote by \textit{N}. The matrix \textit{N} can be parametrized as $N=N^\text{NP}U$, where the nonunitarity is contained in a specific 3$\times$3 triangle matrix \cite{Blennow:2016jkn}
\begin{equation}
N^\text{NP} = 
\left(\begin{array}{ccc}
1-\alpha_{e e} & 0 & 0 \\
-\alpha_{\mu e} & 1-\alpha_{\mu \mu} & 0 \\
-\alpha_{\tau e} & -\alpha_{\tau \mu} & 1-\alpha_{\tau \tau}
\end{array}\right),
\label{Unp}
\end{equation}
where $\alpha_{\ell \ell'} \ll 1$, $\ell, \ell' = e, \mu, \tau$. The current experimental bounds on nonunitarity parameters $\alpha_{l l'}$ are presented in Table \ref{bounds:3}.

\begin{table}[t]
\caption{\label{bounds:3}Current experimental limits of the nonunitarity of the light neutrino mixing matrix \cite{Blennow:2016jkn}. All limits are given at 2$\,\sigma$ confidence limit.}
\begin{center}
\begin{tabular}{|c|c|}\hline
\rule{0pt}{3ex}\textbf{Constraint on} & \textbf{Current bound} \\ \hline
\rule{0pt}{3ex}$\alpha_{e e}$ & 1.3 $\times 10^{-3}$ \\ \hline
\rule{0pt}{3ex}$\alpha_{\mu \mu}$ & 2.2  $\times 10^{-4}$ \\ \hline
\rule{0pt}{3ex}$\alpha_{\tau \tau}$ & 2.8  $\times 10^{-3}$ \\ \hline
\rule{0pt}{3ex}$|\alpha_{\mu e}|$ & 6.8 $\times 10^{-4}$ \\ \hline
\rule{0pt}{3ex}$|\alpha_{\tau e}|$ & 2.7 $\times 10^{-3}$ \\ \hline
\rule{0pt}{3ex}$|\alpha_{\tau \mu}|$ & 1.2 $\times 10^{-3}$ \\ \hline
\end{tabular}
\end{center}
\end{table}

 In the leading order, the matter NSI and nonunitarity parametrizations can be related through the equations \cite{Blennow:2016jkn}:
\begin{align}
\varepsilon_{ee}^m &= -\alpha_{e e} & \varepsilon_{\mu \mu}^m &= \alpha_{\mu \mu}& \varepsilon_{\tau \tau}^m &= \alpha_{\tau \tau} \notag \\
\varepsilon_{e \mu}^m &= \frac{1}{2} \alpha_{\mu e}^* & \varepsilon_{e \tau}^m &= \frac{1}{2} \alpha_{\tau e}^* & \varepsilon_{\mu \tau}^m &= \frac{1}{2} \alpha_{\tau \mu}^*. 
\label{epsarray}
\end{align}

\section{\label{sec:4}Numerical studies}

We shall study what information DUNE could provide us on the parameters $M_\Delta$ and $\lambda_\phi$ through probing the NSI effects on neutrino propagation. Since $\Delta$ is leptophilic, only electron-type matter participates in the NSI effects related to triplet Higgs bosons. In what follows we will use the notation $\varepsilon^{m}_{\alpha\beta} \equiv \varepsilon^{ee}_{\alpha\beta}$. First, we remark that there are limits for both the individual NSI parameters and for their differences
\begin{equation}\label{eq:5a}
\varepsilon_{\alpha\beta}^{\rho\sigma}-\varepsilon_{\alpha'\beta'}^{\rho'\sigma'} = -\frac{M_\Delta^2}{8\sqrt{2}\,G_F\,v^4\,\lambda_\phi^2} \,\, \left( (m_\nu)_{\sigma\beta} \,\, (m_\nu^\dagger)_{\alpha\rho} - (m_\nu)_{\sigma'\beta'} \,\, (m_\nu^\dagger)_{\alpha'\rho'} \right).
\end{equation}
To continue, we consider only matter NSI and rewrite Eq. (\ref{eq:5}) and Eq. (\ref{eq:5a}) in the following forms:
\begin{align}
\frac{M_\Delta^2}{\lambda_\phi^2} &= -\frac{8\sqrt{2}\,G_F\,v^4\,\varepsilon_{\alpha \beta}^{m}}{(m_\nu)_{e \beta} (m_\nu^\dagger)_{\alpha e}} ,\label{eq:num:1} \\
\frac{M_\Delta^2}{\lambda_\phi^2} &= -\frac{8\sqrt{2}\,G_F\,v^4\,(\varepsilon_{\alpha\beta}^{m}-\varepsilon_{\alpha'\beta'}^{m})}{(m_\nu)_{e\beta} \,\, (m_\nu^\dagger)_{\alpha e} - (m_\nu)_{e \beta'} \,\, (m_\nu^\dagger)_{\alpha' e}}\label{eq:num:1a} .
\end{align}
From this expression it is apparent that the upper limits for $|\varepsilon^m_{\alpha\beta}|$ and $|\varepsilon^m_{\alpha\beta}-\varepsilon^m_{\alpha'\beta'}|$ translate to upper limits of $M_\Delta^2/\lambda_\phi^2$. The elements of the light neutrino mass matrix $m_\nu$ are obtained from the equation
\begin{equation}
(m_\nu)^2 = U
\left(
\begin{array}{ccc}
m_1^2 & 0 & 0 \\
0 & m_2^2 & 0\\
0 & 0 & m_3^2
\end{array}
\right) U^{\dagger}
+A\left(
\begin{array}{ccc}
1+\varepsilon_{ee}^m-\varepsilon_{\mu\mu}^m & \varepsilon_{e\mu}^m & \,\,\, \varepsilon_{e\tau}^m \\
\varepsilon_{e\mu}^{m*} & 0 & \,\,\, \varepsilon_{\mu\tau}^m \\
\varepsilon_{e\tau}^{m*} & \varepsilon_{\mu\tau}^{m*} & \,\,\, \varepsilon_{\tau\tau}^m-\varepsilon_{\mu\mu}^m
\end{array}\right).
\label{eq:num:2}
\end{equation}
Note that compared to Eq. (\ref{Vnsi}), we have shifted $\varepsilon^m_{\mu\mu}$ element, akin to \cite{Blennow:2016etl}.  Next, we define a dimensionless quantity,
\begin{equation}\label{eq:C}
C_{\alpha\beta} \equiv 
\begin{dcases}
-\frac{ 8\sqrt{2}\,G_F\,v^4\,\varepsilon_{\alpha \beta}^{m}}{ (m_\nu)_{e \beta} (m_\nu^\dagger)_{\alpha e}},\: \alpha \neq \beta,  \\
-\frac{ 8\sqrt{2}\,G_F\,v^4\,(\varepsilon_{\alpha\beta}^{m}-\varepsilon_{\mu\mu}^{m})}{ (m_\nu)_{e\beta} \,\, (m_\nu^\dagger)_{\alpha e} - (m_\nu)_{e \mu} \,\, (m_\nu^\dagger)_{\mu e}},\: \alpha = \beta.
\end{dcases} 
\end{equation}
This allows us to present the limits in a compact form:
\begin{equation}
\frac{M_\Delta^2}{\lambda_\phi^2} \leq C_{\alpha\beta}.
\end{equation}
 Considering the scales for $G_F$, $v$ and neutrino masses, we expect $C_{\alpha\beta} \gg 1$ and $\lambda_\phi~\ll~M_\Delta$.

 The upper bounds for $M_\Delta/|\lambda_\phi|$ is calculated as follows. Using Eq. (\ref{eq:num:2}), we maximize the denominators in Eq. (\ref{eq:C}) by varying all relevant oscillation parameters within their current experimental bounds. In long baseline neutrino experiments, the electron number density $N_e$ depends on the matter density $\rho$, which is approximately 2700 kg/m$^3$. We take the average beam energy in DUNE to be $E_\nu \approx 2\,\text{GeV}$. Assuming the normal neutrino mass hierarchy, the active neutrino masses $m_2$ and $m_3$ are given by $m_2=\sqrt{m_1^2+\Delta m_{21}^2}$ and $m_3=\sqrt{m_1^2+\Delta m_{31}^2}$ where $m_1 = 0...0.2$~eV, $\Delta m_{21}^2 = 7.50_{-0.17}^{+0.19} \times 10^{-5}$ eV$^2$ and $\Delta m_{31}^2 = 2.524_{-0.040}^{+0.039} \times 10^{-3}$ eV$^2$ (for inverted hierarchy $\Delta m_{32}^2 = -2.514_{-0.041}^{+0.038} \times 10^{-3}$ eV$^2$). The standard oscillation parameters are varied within their 3$\,\sigma$ confidence level (CL) error limits (see \cite{Esteban:2016qun} for exact numbers) to obtain the upper limits of denominators of $C_{\alpha\beta}$. We then take the DUNE 90\% CL upper bounds for the $\varepsilon_{\alpha\beta}^m$ parameters and their differences, and insert the largest possible value of the denominator and the upper limit of $\varepsilon_{\alpha\beta}^m$ into Eq. (\ref{eq:num:1}). This procedure is repeated for the current experimental 90 \% CL upper bounds for matter NSI. The current experimental bounds and the simulated DUNE bounds of these parameters are given in Table \ref{bounds:2}. Comparing these bounds will elucidate the performance and feasability of DUNE as a probe for triplet Higgs model.

\begin{figure}[H]
\begin{center}
\includegraphics[width=\linewidth]{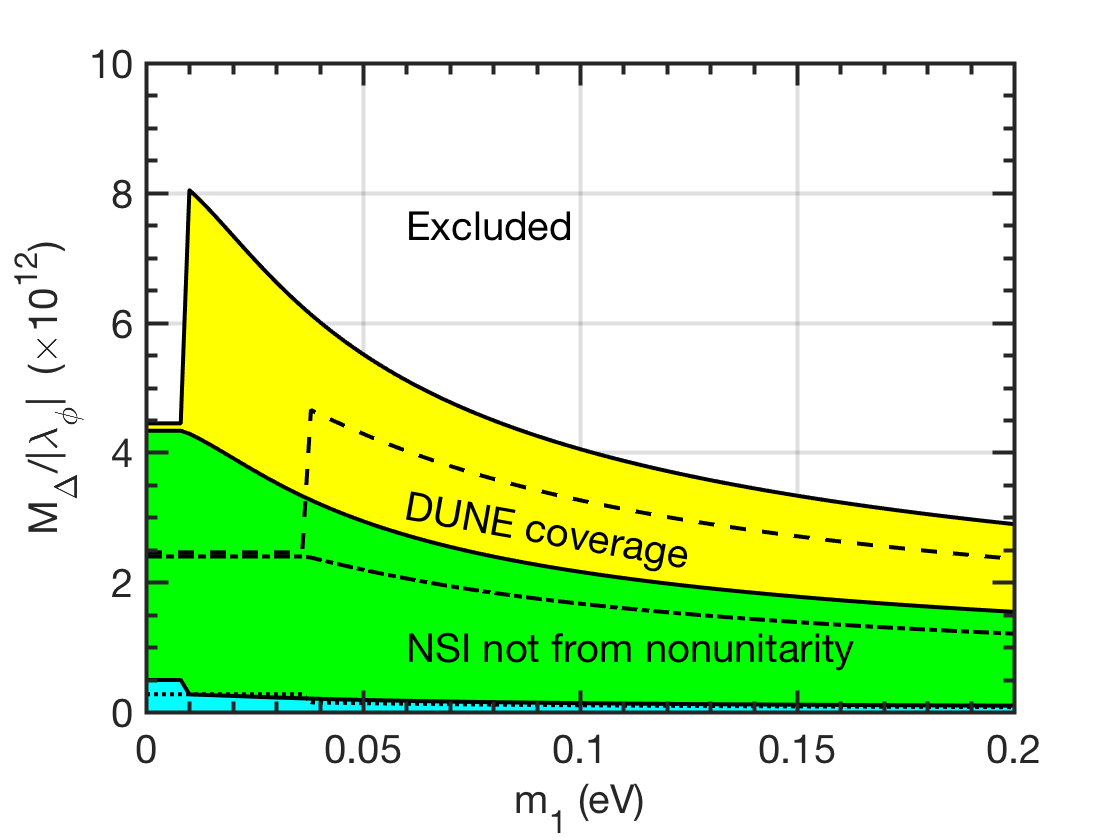}
\end{center}
\caption{\label{fig:2} The allowed values of $M_\Delta/|\lambda_\phi|$ as function of the absolute neutrino mass scale when normal hierarchy is assumed. The white region shows the values of $M_\Delta/|\lambda_\phi|$ which are experimentally excluded at 90\% CL when $m_1$ is the value of the lightest neutrino mass. The yellow region shows the values which are excluded by DUNE. The green region shows the allowed values which can't be constrained by DUNE but which are distinguishable from nonunitarity effects. The blue region shows the allowed values where nonunitarity could in principle be misinterpreted as matter NSI effect. The dashed lines show where these 90\% CL contours for nonunitarity bounds would be, if inverted hierarchy was assumed.}
\end{figure}

 We calculated the expected upper bounds $|C_{ee}|$, $|C_{e\mu}|$, $|C_{e\tau}|$, $|C_{\mu\tau}|$ and $|C_{\tau\tau}|$, for $M_\Delta^2/\lambda_\phi^2$ as function of $m_1$ using the 90\% CL upper bounds for $|\varepsilon_{ee}^m-\varepsilon_{\mu\mu}^m|$, $|\varepsilon_{e\mu}^m|$, $|\varepsilon_{e\tau}^m|$, $|\varepsilon_{\mu\tau}^m|$ and $|\varepsilon_{\tau\tau}^m-\varepsilon_{\mu\mu}^m|$, respectively. Thus, we obtain five $M_\Delta^2/\lambda_\phi^2$ curves as function of $m_1$ and construct the strictest possible upper limit for $M_\Delta^2/\lambda_\phi^2$, namely $C_{\min}(m_1) \equiv \min\left(C_{ee},C_{e\mu},C_{e\tau},C_{\mu\tau},C_{\tau\tau}\right)$. We present this piecewisely combined curve in Fig.~\ref{fig:2}, which was found to be
\begin{align}\label{eq:C1}
C_{\min}^\text{NH}(m_1) &= \begin{cases}
C_{\mu\tau},\: m_1 \lesssim 0.01 \text{ eV}\\
C_{e\mu},\: m_1 \gtrsim 0.01 \text{ eV},
\end{cases}\\
C_{\min}^\text{IH}(m_1) &= \begin{cases}\label{eq:C2}
C_{\mu\tau},\: m_1 \lesssim 0.04 \text{ eV}\\
C_{e\mu},\: m_1 \gtrsim 0.04 \text{ eV},
\end{cases}
\end{align}
where NH and IH label normal and inverted mass hierarchy, respectively.
\begin{table}[H]
\caption{\label{bounds:2}Current experimental upper limits of the matter NSI parameters \cite{Biggio:2009nt} and expected upper bounds after the first run with DUNE \cite{Blennow:2016etl}. All limits are at 90 $\%$ confidence limit.}
\begin{center}
\begin{tabular}{|c|c|c|}\hline
\rule{0pt}{3ex}\textbf{Constraint on} & \textbf{Global bound} &\textbf{ DUNE bound} \\ \hline
\rule{0pt}{3ex}$|\varepsilon^m_{ee}-\varepsilon^m_{\mu\mu}|$ & 4.2 & 0.9 \\ \hline
\rule{0pt}{3ex}$|\varepsilon^m_{e\mu}|$ & 0.3 & 0.074 \\ \hline
\rule{0pt}{3ex}$|\varepsilon^m_{e\tau}|$ & 3.0 & 0.19 \\ \hline
\rule{0pt}{3ex}$|\varepsilon^m_{\mu\tau}|$ & 0.04 & 0.038 \\ \hline
\rule{0pt}{3ex}$|\varepsilon^m_{\tau\tau}-\varepsilon^m_{\mu\mu}|$ & 0.15 & 0.08 \\ \hline
\end{tabular}
\end{center}
\end{table}

 As was pointed out, the nonunitarity effects could be mistaken as matter NSI effects. To estimate how large part these could constitute of the signal, we transpose the current experimental bounds for the nonunitarity parameters $\alpha_{l l'}$ given in Table \ref{bounds:3} into bounds for the NSI parameters $\varepsilon_{\alpha\beta}^ m$ using the relations given in Eq. (\ref{epsarray}). From these bounds we obtained the 90\% CL bound for $M_\Delta/|\lambda_\phi|$ shown in Fig. \ref{fig:2}.

\begin{figure}[H]
\begin{center}
\includegraphics[width=\linewidth]{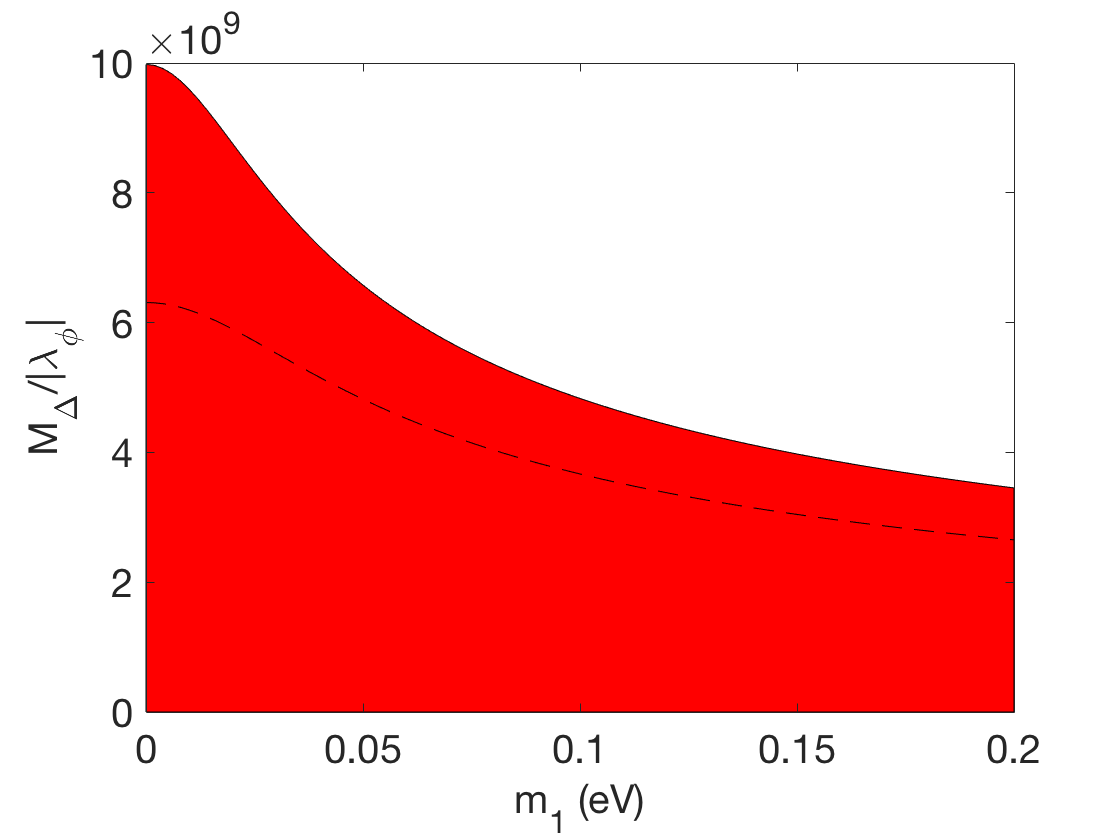}
\end{center}
\caption{\label{fig:3} The allowed values of $M_\Delta/|\lambda_\phi|$ as function of $m_1$ when bounds on the source and detector NSI from CLFV experiments are considered. The white region shows the values of $M_\Delta/|\lambda_\phi|$ which are excluded at 90\% CL, whereas the values in the colored region are still allowed. The dashed line shows the 90\% CL contour in inverted hierarchy.}
\end{figure}

 Similarly, we derive the upper bounds for $M_\Delta/|\lambda_\phi|$ using  the CLFV bounds from Table \ref{bounds:1}. The resulting curve for 90\% CL upper bound for $M_\Delta/|\lambda_\phi|$ is shown in Fig. \ref{fig:3}. One immediately notes that the CLFV constraints are much stricter than the bounds we obtained from the matter NSI, presented in Fig. \ref{fig:2}.

\begin{figure}[H]
\begin{center}
\includegraphics[width=0.75\linewidth]{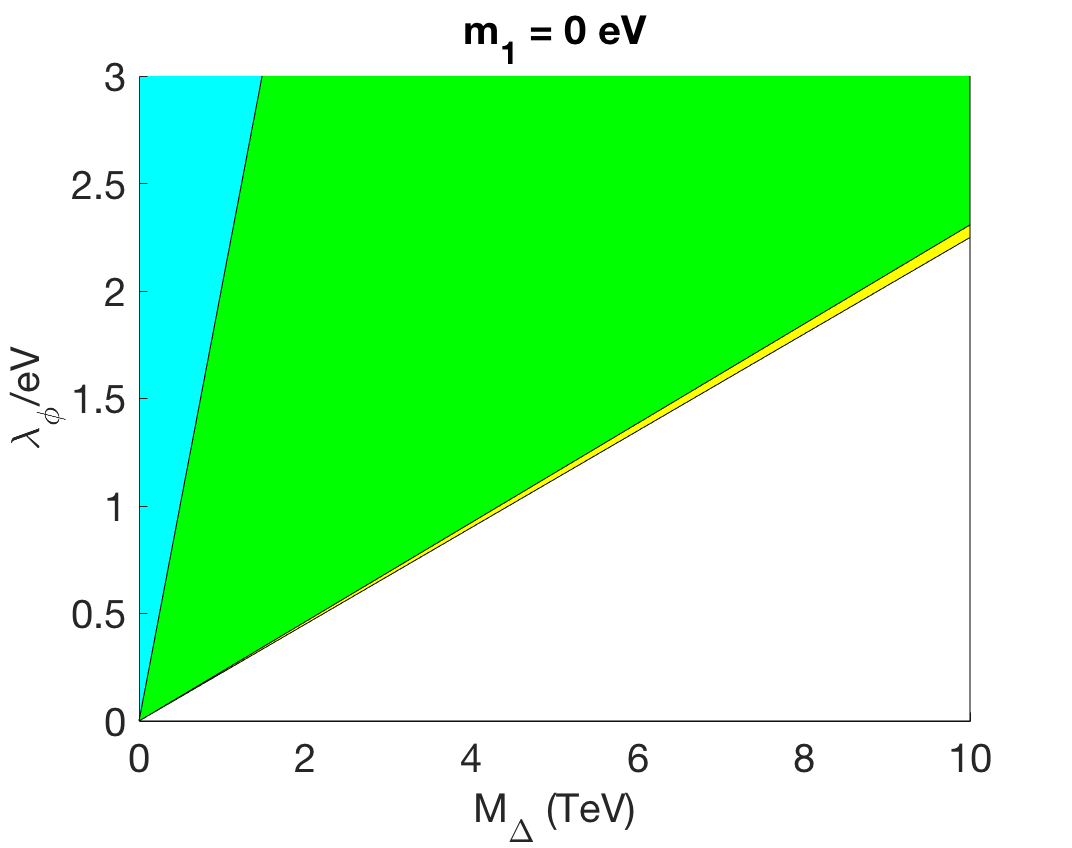}
\includegraphics[width=0.75\linewidth]{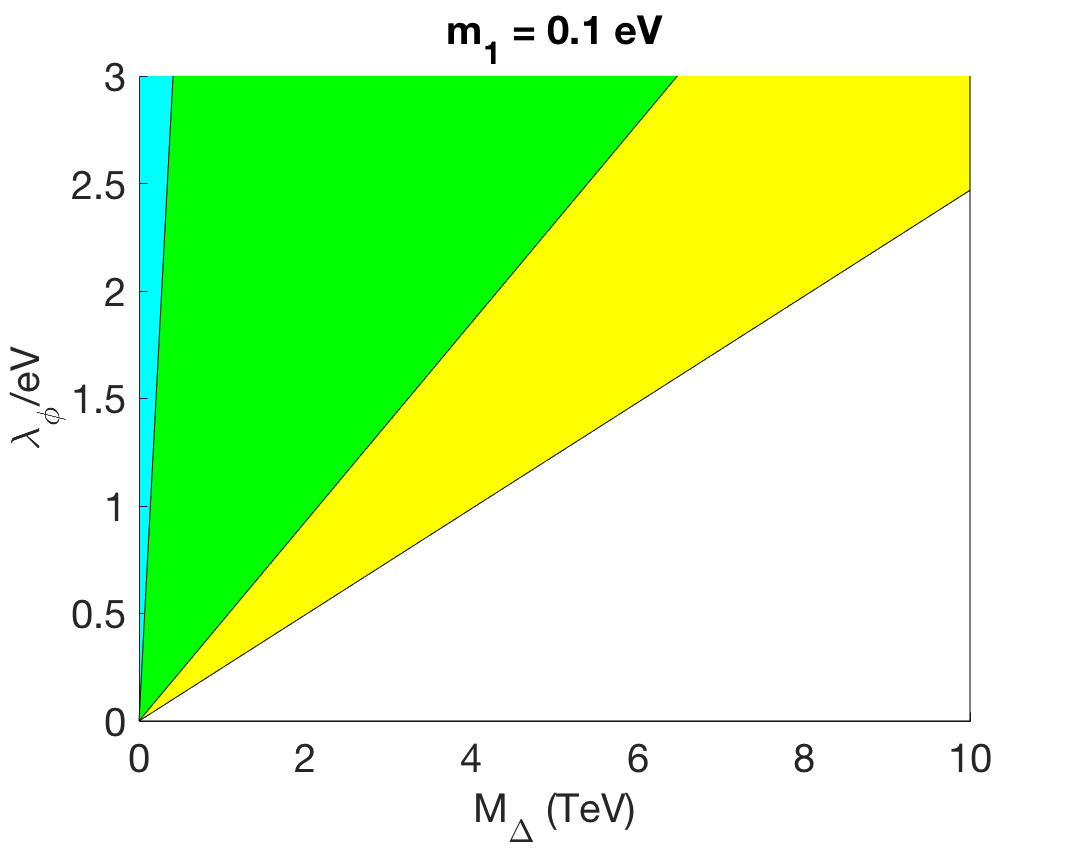}
\end{center}
\caption{\label{fig:4} The allowed values of $M_\Delta$ and $\lambda_\phi$ in $(M_\Delta, \lambda_\phi)$ plane. The colors are chosen as in \ref{fig:2}. We assume lightest neutrino is massless in upper subfigure, and mass of 0.1 eV in lower subfigure.}
\end{figure}

In addition, we present $\lambda_\phi$ as a function of $M_\Delta$ with a constant lightest neutrino mass. We solve \eqref{eq:num:1} for $\lambda_\phi$ and determine the allowed parameter space in $(M_\Delta,\lambda_\phi)$ plane using the aforementioned bounds for matter NSI parameters, considering only bounds arising from current and future long baseline neutrino oscillation experiments. The bounds translate to the conditions
\begin{equation}
\lambda_\phi \geq \frac{M_\Delta}{\sqrt{|C_{\alpha\beta}|}}
\end{equation}
where we again consider the bounds given by $|C_{ee}|$, $|C_{e\mu}|$, $|C_{e\tau}|$, $|C_{\mu\tau}|$ and $|C_{\tau\tau}|$. We pick the tightest constraint, given by $C_{\min}$ from equations \eqref{eq:C1} and \eqref{eq:C2}. The 90 \% CL bound is presented in Fig. \ref{fig:4}.

\begin{figure}[H]
\begin{center}
\includegraphics[width=0.75\linewidth]{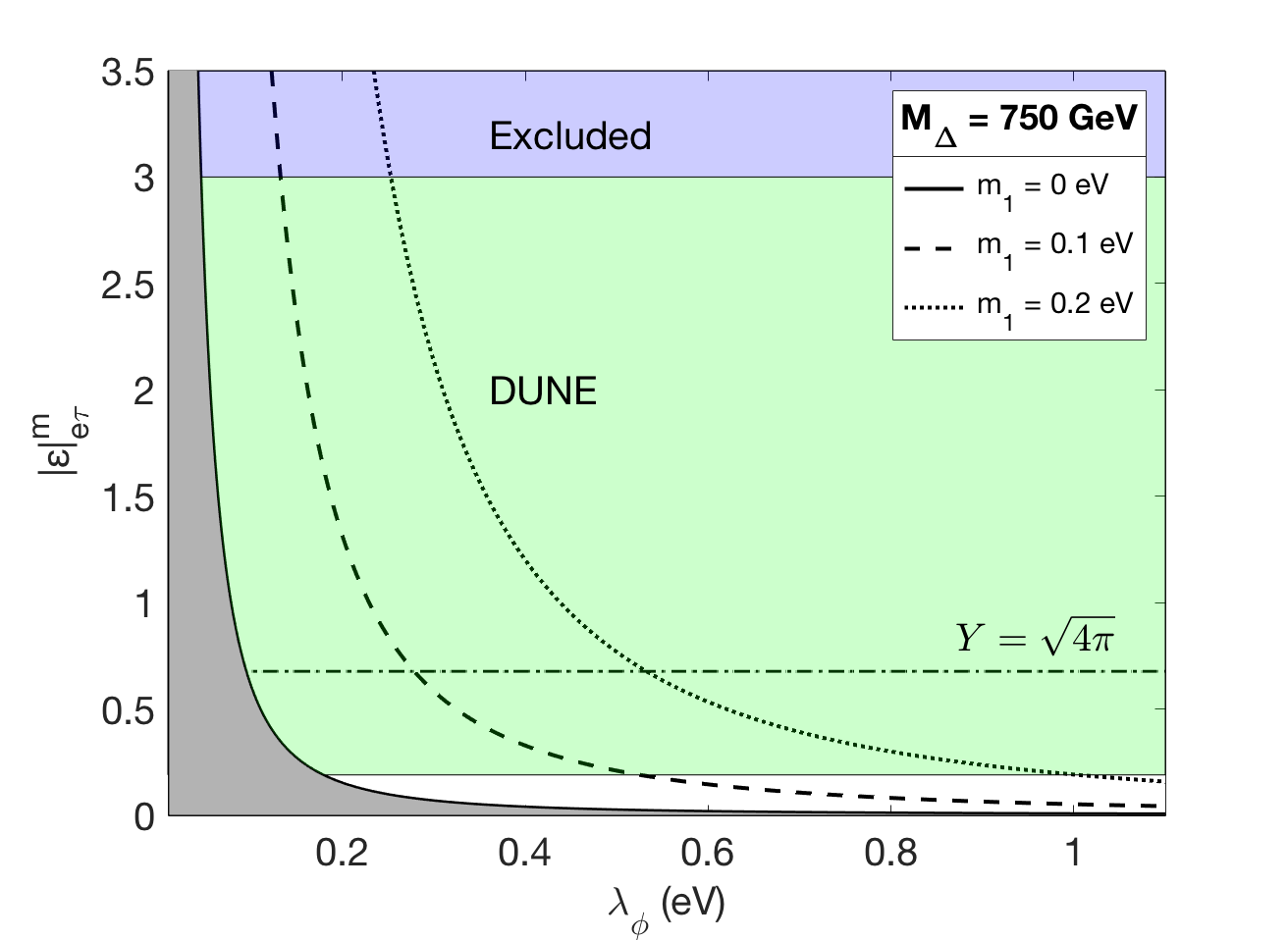}
\includegraphics[width=0.75\linewidth]{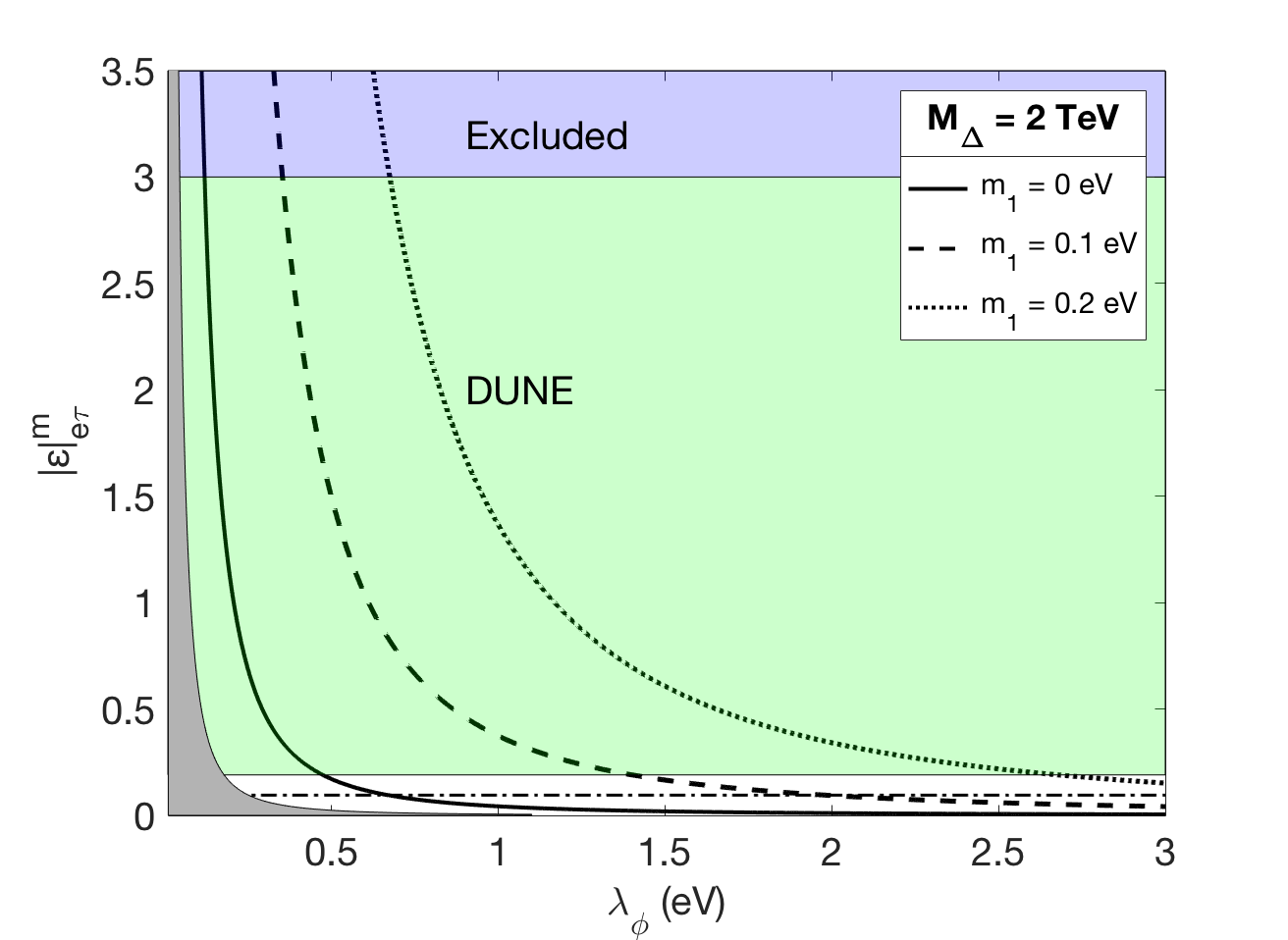}
\end{center}
\caption{\label{fig:5} The allowed values of $|\varepsilon^m_{e\tau}|$ and $\lambda_\phi$ in $(\lambda_\phi, |\varepsilon^m_{e\tau}|)$-plane, assuming normal mass hierarchy. Blue region is excluded by experiments at 90 \% CL. Green region shows the values which are sensitive to DUNE. Gray region is excluded by Higgs searches at LHC, by CMS collaboration \cite{CMS:2017pet}. With a lower limit on degenerate triplet Higgs mass $M_\Delta$ and a fixed lightest neutrino mass, everything below a $m_1$ contour is excluded. This gives us lowest possible value for $\lambda_\phi$ for every $m_1 = 0,\: 0.1,\: 0.2$ eV. At the massless limit we get the absolute limit for $\lambda_\phi$, which may be moderately constrained by DUNE. The Yukawa couplings corresponding to $\varepsilon^m_{e\tau}$ are $<\sqrt{4\pi}$ below the dot-dashed line. In the upper subfigure, the $m_1$ limits are deduced from assumption $M_\Delta=$ 750 GeV and in the lower subfigure, 2 TeV.}
\end{figure}

Since there are known lower limits for $M_\Delta$ \cite{CMS:2017pet}, we may use this information to calculate an experimental lower limit for $\lambda_\phi$. This is done by calculating $|\varepsilon_{ee}^m-\varepsilon_{\mu\mu}^m|$, $|\varepsilon_{e\mu}^m|$, $|\varepsilon_{e\tau}^m|$, $|\varepsilon_{\mu\tau}^m|$ and $|\varepsilon_{\tau\tau}^m-\varepsilon_{\mu\mu}^m|$ as a function of $\lambda_\phi$ from Eq. (\ref{eq:5}) and Eq. (\ref{eq:5a}). We consider $m_1 = 0,\:0.1$ and $0.2$ eV. 
The tightest lower bound for $\lambda_\phi$ is found using $|\varepsilon ^m_{e\tau}|$.
The result is illustrated in Fig.~\ref{fig:5} and in Table~\ref{tab:4}. 

\begin{table}[H]
\caption{\label{tab:4}Current experimental and expected DUNE lower bounds with 90 \% confidence limits for $\lambda_\phi$ inferred from current oscillation parameters, current bounds for NSI and simulated DUNE bounds for NSI. NH and IH correspond to normal and inverse neutrino mass hierarchy, respectively.}
\begin{center}
\begin{tabular}{c|c|c|c|c|}\cline{2-5}
\rule{0pt}{3ex}  & \multicolumn{2}{c}{\textbf{Global} $\lambda_\phi$ (eV)} &  \multicolumn{2}{|c|}{\textbf{DUNE} $\lambda_\phi$ (eV)}\\ \hline
\multicolumn{1}{|c|}{\rule{0pt}{3ex} $m_1$ (eV)} & \textbf{NH} & \textbf{IH} & \textbf{NH} & \textbf{IH} \\ \hline
\multicolumn{1}{|c|}{\rule{0pt}{3ex} 0.0} & 0.031 & 0.045 & 0.120 & 0.178\\ \hline
\multicolumn{1}{|c|}{\rule{0pt}{3ex} 0.1} & 0.129 & 0.133 & 0.509 & 0.526\\ \hline
\multicolumn{1}{|c|}{\rule{0pt}{3ex} 0.2} & 0.251 & 0.253 & 0.997 & 1.006\\ \hline
\end{tabular}
\end{center}
\end{table}

In all cases and results in this section, we have found that in the case the inverted mass hierarchy the bounds are similar as those presented here for the normal hierarchy, just slightly stricter in all cases.


Finally, we conclude this section with remarks on perturbativity. 
By requiring
all the Yukawa couplings to be $\sqrt{4\pi}$ at most, we find
\begin{equation}
M_\Delta \lesssim \frac{617\text{ GeV}}{\sqrt{|\varepsilon_{\alpha\beta}^{\rho\sigma}|}},\hspace{0.5cm} M_\Delta \lesssim \frac{872 \text{ GeV}}{\sqrt{|\varepsilon_{\alpha\beta}^{\rho\sigma}-\varepsilon_{\alpha'\beta'}^{\rho'\sigma'}|}}
\end{equation}
which are acquired by combining Eq.~(\ref{eq:3}) and Eq.~(\ref{eq:4}). Taking into account the lower limits for $M_\Delta$, which we estimate to be approximately 750 GeV \cite{Babu2017, CMS:2017pet}, one can estimate the maximum possible contribution to NSI by the triplet scalar. We find $|\varepsilon_{\alpha\beta}^{\rho\sigma}| \lesssim 0.677$ and $|\varepsilon_{\alpha\beta}^{\rho\sigma}-\varepsilon_{\alpha'\beta'}^{\rho'\sigma'}| \lesssim 1.355$. For CLFV experiments these conditions are fulfilled (see Table~\ref{bounds:1}), but for $|\varepsilon^m_{e\tau}|$ and $|\varepsilon^m_{ee}-\varepsilon^m_{\mu\mu}|$ the experimental limits are not restricting enough (see Fig.~\ref{fig:5}). If DUNE confirms NSI with $|\varepsilon^m_{e\tau}| \gtrsim 0.677$, all of it can't be a manifestation of triplet scalar interactions, and there must be additional new physics contribution. In other words, in the case of confirmation of existence of triplet scalar and large $|\varepsilon^m_{e\tau}|$, this would indicate additional NSI originating from other extensions of SM.


\pagebreak
\thispagestyle{empty} 

\section{\label{sec:5}Summary and conclusions}

The triplet Higgs bosons  $(\Delta=\Delta^{++},\Delta^{+},\Delta^{0})$ utilized in the Type-II seesaw model will affect neutrino propagation in matter by mediating interactions not present in the Standard Model. We have studied in this work how the sensitivity of DUNE for NSI interactions can be utilized to derive a constraint for the quantity $M_\Delta/|\lambda_\phi|$, where $M_\Delta$ is the mass of the triplet bosons and $\lambda_\phi$ the strength of the coupling $\Delta\phi\phi$ between the triplet Higgs and the standard Higgs doublet. We found that a long baseline experiment with the specifications of the proposed DUNE can reach the upper bound  $M_\Delta/|\lambda_\phi|\lesssim (3 $~---~$4) \times 10^{12}$ at 90\% CL. If the ratio $M_\Delta/|\lambda_\phi|$ were above this value, the effects of the NSI neutrino-matter interactions caused by the triplet boson exchange would be seen in the DUNE oscillation data. We found the bound to be sensitive of the neutrino mass ordering (normal or inverted).
 
 We found that in long baseline experiments the strictest bound on $M_\Delta/|\lambda_\phi|$ arises from the $|\varepsilon_{\mu\tau}^m|$ and $|\varepsilon_{e\mu}^m|$ constraints. DUNE would be able to improve the upper limits of these NSI parameters, as indicated in Table~\ref{bounds:2}. The sensitivity to the matter NSI parameters has been previously studied for the proposed HyperKamiokande and T2HKK experiments, and the upper limit on $|\varepsilon_{e e}^m - \varepsilon_{\mu \mu}^m|$ achievable in these experiments is estimated to be improved, see
\cite{Fukasawa:2016lew}. 
 
 In long baseline experiments nonunitarity of the mixing matrix of the ordinary light neutrinos might give  similar effects on the oscillation probabilities than the NSI due to the triplet Higgses, which might lead to a misinterpretation of the data. We found out that the effects caused by nonunitarity depend strongly on the lightest neutrino mass $m_1$ and are well below the sensitivity of DUNE. Only when $m_1\ll 1$ eV, the signals interpreted as triplet-Higgs NSI effects could actually be caused by nonunitarity of the neutrino mixing. 
 
 We also showed that the existing constraints on the various NSI parameters obtained by studying low-energy charged lepton flavour violation (CLFV) processes, such as  like $\mu^- \to e^-e^+e^-$, correspond, when interpreted in terms of the triplet Higgs model, to constraints on $M_\Delta/|\lambda_\phi|$, which are for all values of the lightest neutrino mass orders of magnitude more stringent than one would reach in DUNE.
 
To summarize, the study of NSI interactions in long-baseline neutrino-oscillation experiments, considered in this work, and at low-energy processes would give crucially central information on the triplet-Higgs model and Type-II seesaw model, in particular when combined with the information one can get in collider experiments for the lower limit of  the triplet mass $M_\Delta$ (see e.g. Ref. \cite{CMS:2017pet}). For normal mass hierarchy and a triplet mass $M_\Delta = 750$\text{ GeV}, for example, the future DUNE data will indicate $|\lambda_\phi| \gtrsim 0.120\,\text{ eV}$, a clear improvement of the strictest current limit we calculated in Table \ref{tab:4}, $|\lambda_\phi| \gtrsim 0.031\,\text{ eV}$.

\section*{Acknowledgements}
KH acknowledges the H2020-MSCA-RICE-2014 grant no. 645722 (NonMinimalHiggs). TK expresses his gratitude to the Magnus Ehrnrooth foundation for financial support. SV thanks University of Jyv\"askyl\"a for funding a research visit to Centro de F\'isica Te\'orica de Part\'iculas, University of Lisbon, where part of this work was done.

\bibliography{bibliography} 
\end{document}